\documentclass{emulateapj}

\newcommand{\LO}{\ensuremath{L_{\odot}}}   
\newcommand{\MO}{\ensuremath{M_{\odot}}}   
\newcommand{\ZO}{\ensuremath{Z_{\odot}}}   

\slugcomment{Accepted for Publication in the Astrophysical Journal}

\shorttitle{Nature of SXDF~850.6}

\shortauthors{Hatsukade et al.}

\begin{document}

\title{Unveiling the Nature of Submillimeter Galaxy SXDF~850.6}

\author{
		B. Hatsukade\altaffilmark{1},
		D. Iono\altaffilmark{2},
		T. Yoshikawa\altaffilmark{3}
		M. Akiyama\altaffilmark{3},
		J.\,S.\ Dunlop\altaffilmark{4},
		R.\,J.\ Ivison\altaffilmark{5,4},
		A. B. Peck\altaffilmark{6},
		S.~Ikarashi\altaffilmark{1},
		A.~Biggs\altaffilmark{7}, 
 		H. Ezawa\altaffilmark{8}, 
 		H. Hanami\altaffilmark{9},
 		P. Ho\altaffilmark{10}, 
 		D. H. Hughes\altaffilmark{11}, 
 		R. Kawabe\altaffilmark{2}, 
 		K. Kohno\altaffilmark{1,12}, 
		S.~Matsushita\altaffilmark{10}, 
 		K.~Nakanishi\altaffilmark{2}, 
 		N.~Padilla\altaffilmark{13},
 		G. Petitpas\altaffilmark{14}, 
		Y. Tamura\altaffilmark{2}, 
		J. Wagg\altaffilmark{11,15,16}, 
		D. J. Wilner\altaffilmark{16},
 		G. W. Wilson\altaffilmark{17}, 
		T.~Yamada\altaffilmark{3}, 
 		and~M.~S.~Yun\altaffilmark{17}
}

\altaffiltext{1}{Institute of Astronomy, the University of Tokyo, 2-21-1 Osawa, Mitaka, Tokyo 181-0015, Japan}
\email{hatsukade@ioa.s.u-tokyo.ac.jp}
\altaffiltext{2}{Nobeyama Radio Observatory, Minamimaki, Minamisaku, Nagano 384-1805, Japan}
\altaffiltext{3}{Astronomical Institute, Tohoku University, Aramaki, Aoba-ku, Sendai, Miyagi 980-8578, Japan}
\altaffiltext{4}{Scottish Universities Physics Alliance, Institute for Astronomy, School of Physics and Astronomy, University of Edinburgh, Royal Observatory, Edinburgh EH9 3HJ, UK}
\altaffiltext{5}{UK Astronomy Technology Centre, Science and Technology Research Council, Royal Observatory, Blackford Hill, Edinburgh EH9 3HJ, UK}
\altaffiltext{6}{Joint ALMA Observatory, Avenida El Golf 40, Piso 18, Las Condes 7550108 Santiago, Chile}
\altaffiltext{7}{European Southern Observatory, Karl-Schwarzschild-Stra\ss e 2, D-85748 Garching, Germany}
\altaffiltext{8}{National Astronomical Observatory of Japan, 2-21-1 Osawa, Mitaka, Tokyo 181-8588, Japan}
\altaffiltext{9}{Physics Section, Faculty of Humanities and Social Sciences, Iwate University, Morioka 020-8550, Japan}
\altaffiltext{10}{Academia Sinica Institute of Astronomy and Astrophysics, P.O. Box 23-141, Taipei 10617, Taiwan}
\altaffiltext{11}{Instituto Nacional de Astrofisica, \'{O}ptica y Electr\'{o}nica (INAOE), Aptdo. Postal 51 y 216, 72000 Puebla, Pue., Mexico}
\altaffiltext{12}{Research Center for the Early Universe, University of Tokyo, 7-3-1 Hongo, Bunkyo, Tokyo 113-0033, Japan}
\altaffiltext{13}{Departamento de Astronom\'{\i}a y Astrof\'{\i}sica, Pontificia Universidad Cat\'{o}lica de Chile, Vicu\~{n}a Mackenna 4860, Santiago, Chile}
\altaffiltext{14}{Harvard-Smithsonian Center for Astrophysics, Submillimeter Array, 645 North A`ohoku Place, Hilo, HI 96720, USA}
\altaffiltext{15}{European Southern Observatory, Alonso de C\'{o}rdova 3107, Vitacura, Casilla 19001, Santiago 19, Chile}
\altaffiltext{16}{Harvard-Smithsonian Center for Astrophysics, 60 Garden Street, Cambridge, MA 02138, USA}
\altaffiltext{17}{University of Massachusetts, Department of Astronomy, Amherst, MA01003, USA}

\begin{abstract}
We present an 880~$\micron$ Submillimeter Array (SMA) detection of the submillimeter galaxy SXDF~850.6. 
SXDF~850.6 is a bright source ($S_{\rm 850\,\mu m} = 8$~mJy) detected in the SCUBA Half Degree Extragalactic Survey (SHADES), 
and has multiple possible radio counterparts in its deep radio image obtained at the VLA. 
Our new SMA detection finds that the submm emission coincides with the brightest radio emission that is found $\sim$$8''$ north of the coordinates determined from SCUBA. 
Despite the lack of detectable counterparts in deep UV/optical images, we find a source at the SMA position in near-infrared and longer wavelength images. 
We perform SED model fits to UV--optical--IR photometry ($u, B, V, R, i', z', J, H, K$, 3.6~$\micron$, 4.5~$\micron$, 5.8~$\micron$, and 8.0~$\micron$) 
and to submm--radio photometry (850~$\micron$, 880~$\micron$, 1100~$\micron$, and 21~cm) independently, 
and we find both are well described by starburst templates at a redshift of $z \simeq 2.2 \pm 0.3$. 
The best-fit parameters from the UV--optical--IR SED fit are 
a redshift of $z = 1.87^{+0.15}_{-0.07}$, 
a stellar mass of $M_{\star} = 2.5^{+2.2}_{-0.3} \times 10^{11}\ \MO$, 
an extinction of $A_V = 3.0^{+0.3}_{-1.0}$~mag, 
and an age of $720^{+1880}_{-210}$~Myr. 
The submm--radio SED fit provides a consistent redshift of $z \sim 1.8$--2.5, an IR luminosity of $L_{\rm IR}$ = (7--26)~$\times 10^{12}\ \LO$, and a star formation rate of 1300--4500~$\MO\ \rm{yr}^{-1}$. 
These results suggest that SXDF~850.6 is a mature system already having a massive amount of old stellar population constructed before its submm bright phase and is experiencing a dusty starburst, possibly induced by major mergers. 

\end{abstract}


\section{Introduction}

Millimeter/submillimeter surveys have revolutionized observational cosmology by uncovering a substantial new population of mm/submm-bright dusty starburst galaxies at high redshifts (SMGs) \citep[e.g.,][and see also \citealt{blai02} for a review]{smai97, barg98, hugh98, scot02, copp06, grev04, laur05, scot08, perera08, aust09, tamu09}. 
The energy source of mm/submm emission is primarily from intense star formation activity, with star formation rates (SFRs) of 100~$\MO\ \rm{yr}^{-1}$ to several 1000~$\MO\ \rm{yr}^{-1}$, and possibly partially from an active galactic nucleus (AGN) \citep[e.g.,][]{alex05}. 
While spectroscopic observations of radio-identified SMGs find a median redshift of 2.2 \citep{chap05}, several SMGs have now been found up to and beyond $z=4$ \citep[e.g.,][]{capa08, copp09, dadd09}. 
There is a hypothesis that SMGs are progenitors of present-day massive ellipticals \citep[e.g.,][]{lill99, smai04}, however, little is known about their evolution process. 

While multi-wavelength analysis is essential in order to understand the nature of SMGs, the coarse angular resolution of single dish telescopes prevents a precise determination of the exact optical/NIR counterparts. 
One of the most successful ways to pinpoint the location of the submm emission is to obtain high resolution, deep radio imaging \citep[e.g,][]{ivis98, ivis00, ivis02, smai00, barg00}. 
Although this technique reveals robust radio counterparts of $\sim$50\%--80\% of submm sources \citep[e.g.,][]{ivis05, ivis07, wagg09}, sometimes multiple radio counterpart candidates are found for a source. 

The most accurate means of achieving high precision astrometry on the submillimeter emission is clearly to observe with high angular resolution at the wavelength of the original detection.
In this respect, the Submillimeter Array \citep[SMA;][]{ho04} has proved to be a powerful instrument \citep{iono06a, iono06b, wang07, youn07, youn08, ivis08}.

Here we present the results from the SMA observations toward an 8~mJy submm source, SXDF~850.6, detected in the SCUBA Half Degree Extragalactic Survey (SHADES). 
SHADES has observed a large area of the sky (720 arcmin$^2$) with high sensitivity ($1\sigma \sim$~2~mJy) with the purpose of obtaining a statistically significant unbiased sample of submillimeter sources \citep{mort05, copp06}. 
The regions covered by SHADES are divided between two fields, the Lockman Hole and the Subaru/\textit{XMM-Newton} Deep Field (SXDF). 
SXDF~850.6 is a source in the SXDF with multiple optical, IR, and radio counterparts \citep{ivis07, clem08}, but no established submm source identification. 
The strongest radio emission has no confirmed optical counterpart, but the two secondary radio peaks both have apparent optical associations. 

\S~2 outlines the observational and calibration details, and the results are presented in \S~3. 
In \S~4, multi-wavelength data are described. 
The results of SED model fitting using the photometry are presented in \S~5. 
In \S~6, we discuss the nature of SXDF~850.6. 
A summary is presented in \S~7. 
Throughout the paper, magnitudes are in the AB system, and we adopt a cosmology with $H_0=70$ km s$^{-1}$ Mpc$^{-1}$, $\Omega_{\rm{M}}=0.3$, and $\Omega_{\Lambda}=0.7$.

\section{Observations and Data Reduction}

SXDF~850.6 was observed on September 21 and October 7, 2004, and on October 9 and 14, 2005, using a compact configuration with 7 -- 8 antennas of the SMA. 
The phase center was positioned at the submm source centroid which is $\alpha$(J2000)~=~$02^h$~$17^m$~$29.80^s$ and $\delta$(J2000)~=~$-05^{\circ}$~$03'$~$26''.00$. 
The unprojected baseline length ranged from 23~m to 139m. 
The SMA correlator was equipped with 2~GHz total bandwidth in each sideband, yielding a total of 4~GHz bandwidth for continuum observations. 
A continuum channel was generated by vector averaging all of the channels after calibration. 
The SIS receivers were tuned to 345~GHz for the USB, yielding 335~GHz for the LSB. 
Interferometric pointing was checked at the beginning of the track and the pointing offsets were usually within $\sim \pm 5''$ (15\% of the primary beam) for all antennas. 
We used an integration time of 30 seconds.

The raw SMA data was calibrated using the MIR package \citep{scov93}. 
Passband calibration was done using bright QSOs and planets observed during the track. 
Antenna based phase calibration was done using J0238+166~(1.02~Jy), J0423-013~(1.67~Jy) and J0132-169~(0.84~Jy). 
The flux levels of all sources were normalized using the the quasar flux estimates derived from the primary flux standard Uranus. 
Imaging was carried out in MIRIAD \citep{saul95}.  
Maximum sensitivity was achieved by adopting natural weighting, which gave a final synthesized beam size of $2\farcs32 \times 2\farcs19$ (P.A. = $79.1^{\circ}$) and an RMS noise of 1.2~mJy.
Because the source is close to zero in declination, the uv coverage is undersampled in the north-south direction and the resultant synthesized beam has multiple sidelobes at the 30\% level along the north-south direction about $14''$ away from the synthesized beam. 
The final map was made after adding the data from all four days and two sidebands, and corrected for the attenuation by the $35''$ primary beam of the SMA. 
The astrometric accuracy is likely dominated by the low S/N of the image, and we assess $\sim$$0\farcs4$.

\section{Results}

\begin{figure}
\epsscale{1.15}
\plotone{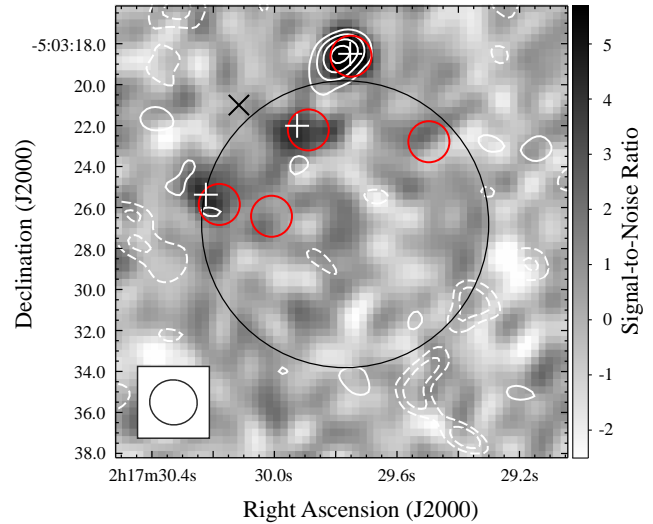}
\caption{
The SMA contours overlaid on the VLA 21~cm image. 
The contours represent $-3$, $-2$, 2, 3, 4, and $5 \sigma$ (where $1 \sigma = 1.2$~mJy). 
The synthesized beam ($2\farcs32 \times 2\farcs19$) is shown in lower left. 
The dark circle represents the approximate size of the SCUBA/JCMT beam (FWHM $\sim$ $14''$).
The dark cross, white crosses, and red circles are the AzTEC position (Ikarashi et al. in prep.), three radio counterpart candidates in \citet{ivis07}, and MIPS 24~$\micron$ sources in the SWIRE catalog, respectively. 
}
\label{fig:map}
\end{figure}

The synthesized map is shown in Figure~\ref{fig:map}. 
A source was detected at about 6 sigma significance $8\farcs2$ north of the SCUBA coordinates. 
The source appears slightly elongated in the northeast-southwest direction but this is likely caused by the low S/N, and the source is entirely unresolved with the SMA beam. 
Based on a point source fit to the visibilities, the derived flux is $6.9 \pm 1.2$~mJy, and the coordinates are $\alpha$(J2000)~=~$02^h$~$17^m$~$29.79^s$ and $\delta$(J2000)~=~$-05^{\circ}$~$03'$~$18\farcs65$. 
The derived flux is consistent with the SCUBA 850~$\micron$ flux of $8.15 \pm 2.2$~mJy \citep{copp06} within the uncertainties of both measurements.

\section{Multi-wavelength Data}

\begin{figure*}
\epsscale{0.7}
\plotone{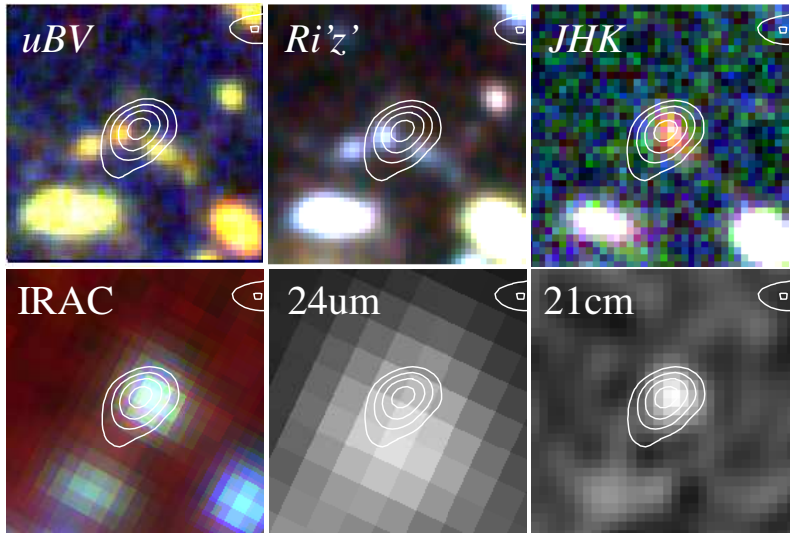}
\caption{
  	Multi-wavelength images of SXDF~850.6 with the SMA contours (2, 3, 4, and 5$\sigma$). 
	The size of each image is $10'' \times 10''$ and north is up. 
  	From left to right: 
  	rgb image of MOSAIC~II/$u$, SuprimeCam/$B$, and $V$; 
  	rgb image of SuprimeCam/$R$, $i'$, and $z'$; 
  	rgb image of WFCAM/$J$, $H$, and $K$; 
  	rgb image of IRAC/ch1 (blue), ch2 (green), ch3 and 4 (red); 
  	MIPS 24~$\micron$; 
  	VLA 21~cm. 
    }
\label{fig:stamp}
\end{figure*}

Multi-wavelength images around SXDF~850.6 are shown in Figure~\ref{fig:stamp}. 
While an obvious optical counterpart is not seen in the images from $u$ to $z'$ bands, the NIR images represented in $JHK$ and longer wavelength images show a presence of a counterpart. 
The photometry of the counterpart and upper limits at the SMA position are presented in Table~\ref{tab:flux}. 

\subsection{Radio and Millimeter}
\citet{ivis07} identify three radio counterpart candidates for SXDF~850.6 in their deep VLA 21~cm map based on the probability analysis method of \cite{down86}. 
All three radio sources have corresponding MIPS 24~$\micron$ sources \citep{clem08}, as shown in Figure~\ref{fig:map}. 
The SMA observation reveals that the correct radio counterpart is the most distant source from the SCUBA centroid, and the brightest in 24~$\micron$ emission. 
This demonstrates the effectiveness of the SMA in identifying SMG counterparts, especially in situations with multiple counterpart candidates. 

The radio coordinates derived from reanalysis of the VLA image (Arumugam et al. in prep.) are $\alpha$~(J2000)~=~$02^h$~$17^m$~$29.755^s$ ($\pm 0.005^s$) and $\delta$~(J2000)~=~$-05^{\circ}$~$03'$~$18\farcs40$ ($\pm 0\farcs08$), with peak and integrated flux of $99.8~\mu$Jy and $100.0 \pm 10.6~\mu$Jy respectively. 
The close similarity of the peak and integrated flux suggests that the radio emission is unresolved with the $1\farcs79 \times 1\farcs51$ beam (P.A. = $6.1^\circ$). 

SXDF~850.6 is detected at 1100 $\micron$ with deboosted flux of $3.9 \pm 0.5$~mJy (Ikarashi et al. in prep.) using the AzTEC camera \citep{wils08} on the Atacama Submillimeter Telescope Experiment \citep[ASTE;][]{ezaw04, ezaw08}. 
The AzTEC coordinates are consistent with the SMA position within a $2\sigma$ error circle.

\subsection{Mid-IR}\label{sec:ir}

{\sl Spitzer}/IRAC and MIPS data are taken from the SpUDS archive. 
The 3$\sigma$ detection limits are 0.58, 0.89, 5.7, and 5.3 $\mu$Jy in IRAC bands (3.6, 4.5, 5.8, and 8.0 $\micron$), 
and 36~$\mu$Jy in MIPS 24~$\micron$ band. 
The photometry was performed using the SExtractor package \citep{bert96}. 
The photometry in IRAC bands is conducted with $3\farcs8$ diameter aperture, and 
the photometric zero-points of the images are determined using colors of early-type stars between $K$-band and IRAC bands following \cite{lacy05}. 
In order to derive total flux, the aperture photometry are corrected by assuming IRAC PSFs by a factor of 1.36, 1.40, 1.65, and 1.84,  respectively, 
The photometry in the 24~$\micron$ band is conducted using 12$''$ diameter aperture. 
The aperture magnitude is corrected by a factor of 1.70. 

We find emission at the SMA position in all IRAC bands and MIPS 24~$\micron$ band. 
Note that the 24~$\micron$ flux in Table~\ref{tab:flux} should be used as an upper limit since sources around the SMA position are blended in the 24~$\micron$ image (Figure \ref{fig:stamp}).

\subsection{$JHK$}\label{sec:jhk}

J, H, and K band images are obtained from UKIDSS (UKIRT InfraRed Deep Sky Surveys) Ultra Deep Survey Third Data Release \citep{laur05}. 
The 3$\sigma$ detection limits are 24.28, 24.08, and 24.18 mag, respectively.  

A faint source is detected at the SMA position in all bands. 
We perform photometry using the SExtractor. 
Aperture photometry with $1\farcs8$ diameter aperture in the $J$, $H$, and $K$ bands are corrected by difference between the aperture and the total (MAG\_AUTO) magnitudes in the $K$-band.

\subsection{$uBVRi'z'$}\label{sec:ubvriz}
We present a $u$-band image of MOSAIC~II on CTIO 4-m telescope (Fujishiro et al. in prep.), $B, V, R, i'$, and $z'$ images of SuprimeCam on Subaru telescope in Figure \ref{fig:stamp}. 
$B$, $V$, $R$, $i'$, and $z'$ band images are obtained from Subaru/XMM-Newton Deep Survey database \citep[SXDS;][]{furu08}. 
No galaxy is detected at the SMA position in these bands. 
The 3$\sigma$ detection limits are 26.00, 27.94, 27.68, 27.43, 27.45, and 26.43 mag, respectively. 

Near the SMA position, three galaxies are seen; $\sim$$2''$ south-east, $\sim$$0\farcs8$ south-east, and $\sim$$1''$ south-west to the SMA position. 
We derive photometric redshifts for the sources and find that they are likely to be at $z < 1$, and it is unlikely that they are related to the SMA source (see \S~\ref{sec:fit}).

\begin{deluxetable}{cccc}
\tabletypesize{\scriptsize}
\tablewidth{0pt}
\tablecaption{Fluxes of SXDF~850.6\label{tab:flux}}
\tablehead{
\colhead{Band} & \colhead{Flux} & \colhead{Instrument} & \colhead{Ref.} 
} 
\startdata
$u$           & $<$$0.0148$ $\mu$Jy & MOSAIC~II & 1 \\
$B$           & $<$$0.0248$ $\mu$Jy & SuprimeCam & 1 \\
$V$           & $<$$0.0315$ $\mu$Jy & SuprimeCam & 1 \\
$R$           & $<$$0.0397$ $\mu$Jy & SuprimeCam & 1 \\
$i'$          & $<$$0.0389$ $\mu$Jy & SuprimeCam & 1 \\
$z'$          & $<$$0.0996$ $\mu$Jy & SuprimeCam & 1 \\
$J$           & $1.40^{+0.23}_{-0.20}$ $\mu$Jy & WFCAM & 1 \\
$H$           & $3.21^{+0.23}_{-0.22}$ $\mu$Jy & WFCAM & 1 \\
$K$           & $9.17 \pm 0.23$ $\mu$Jy & WFCAM & 1 \\
3.6 $\micron$ & $29.0 \pm 0.8$ $\mu$Jy & IRAC & 1 \\
4.5 $\micron$ & $39.3 \pm 1.1$ $\mu$Jy & IRAC & 1 \\
5.8 $\micron$ & $54.3 \pm 1.9$ $\mu$Jy & IRAC & 1 \\
8.0 $\micron$ & $38.2 \pm 1.4$ $\mu$Jy & IRAC & 1 \\
24  $\micron$ & $630  \pm 30 $ $\mu$Jy & MIPS & 1 \\
70  $\micron$ & $<$30 mJy & MIPS & 2 \\
160 $\micron$ & $<$200 mJy & MIPS & 2 \\
450 $\micron$ & $<$81 mJy & SCUBA & 3 \\
850 $\micron$ & $8.15 \pm 2.2$ mJy & SCUBA & 3 \\
880 $\micron$ & $6.9 \pm 1.2 $ mJy & SMA & 1 \\
1100 $\micron$& $3.9 \pm 0.5$ mJy & AzTEC & 4 \\
21 cm         & $100.0 \pm 10.6$ $\mu$Jy & VLA & 5 \\
0.2--12 keV   & $<$$6.3 \times 10^{-15}$ erg cm$^{-2}$ s$^{-1}$ & {\sl XMM-Newton} & 6 \\
\enddata
\tablecomments{
Limits are 3$\sigma$. 
}
\tablerefs{
(1) This work; 
(2) Flux limits in the SWIRE catalog (Surace et al. in prep.); 
(3) Coppin et al. 2006; 
(4) Ikarashi et al. in prep.; 
(5) Arumugam et al. in prep.
(6) FLIX: upper limit server for XMM-Newton data provided by the XMM-Newton Survey Science Centre
}
\end{deluxetable}

\section{SED Fitting}\label{sec:fit}

We perform SED model fitting to the photometry data from UV to radio to estimate photometric redshifts and other physical properties (stellar mass, $V$-band extinction, star formation timescale, and metallicity). 
The photometric data are divided into two wavelength ranges: 
(1) from UV to 8.0~$\micron$, where stellar emission is dominant, 
(2) from submm to radio, where thermal dust and synchrotron emission is dominant.

\subsection{UV -- Optical -- IR}\label{sec:stellar_sed}

\begin{figure}
\epsscale{1.15}
\plotone{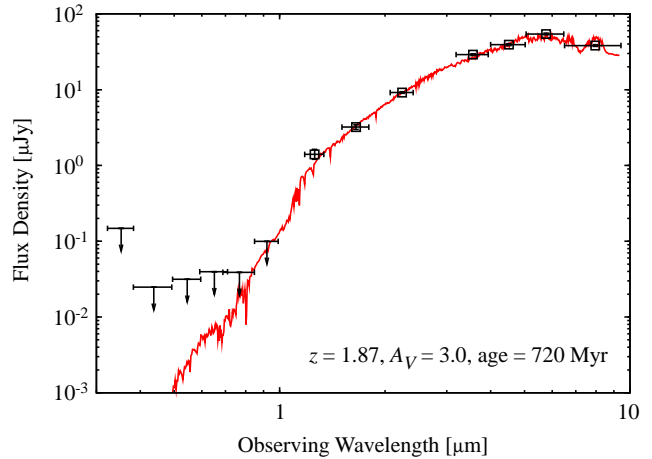}
\caption{
Best-fit SED obtained in UV--optical--IR SED fit along with photometry data. 
Downward arrows represent 3$\sigma$ upper limits. 
}
\label{fig:stellar_sed}
\end{figure}

SED fits are performed to the photometry ranging from UV to 8.0~$\micron$ ($u, B, V, R, i', z', J, H, K$, 3.6~$\micron$, 4.5~$\micron$, 5.8~$\micron$, and 8.0~$\micron$) using the Hyperz code \citep{bolz00}. 
We use a synthetic spectral library of GALAXEV \citep{bruz03} for SED templates. 
We adopt the Calzetti extinction law \citep{calz00} and the \cite{chab03} initial mass function with lower and upper cutoff mass of 0.1 and 100~\MO. 
Two star formation histories are assumed: 
(i) exponentially decaying SFR with a timescale of $\tau$ (i.e., ${\rm SFR} \propto e^{-t/\tau}$), 
(ii) constant SFR. 
We calculate $\chi^2$ values for each SED template with free parameters of redshift ($z = 0$--6), stellar mass ($M_{\star}$), $V$-band attenuation ($A_V = 0.0$--6.0), star formation timescale ($\tau = 0.01$--5~Gyr), and metallicity ($Z = 1.0, 0.4, 0.2$~\ZO). 
The best-fit results are obtained with the exponentially decaying SFR, 
$z = 1.87^{+0.15}_{-0.07}$, 
$M_{\star} = 2.5^{+2.2}_{-0.3} \times 10^{11}\ \MO$, 
$A_V = 3.0^{+0.3}_{-1.0}$~mag, 
and age = $720^{+1880}_{-210}$~Myr 
(Table~\ref{tab:fit} and Figure \ref{fig:stellar_sed}). 
The results suggest that SXDF~850.6 has a large population of old stars, with large amount of dust obscuring the stellar emission. 
The stellar mass is comparable to those of massive star-forming galaxies at $z \sim 2$ \citep[e.g.,][]{shap05, erb06a} and consistent with stellar masses recently derived for other SMGs ($\sim$$10^{11}$--$10^{12}$~\MO) \citep[e.g.,][]{dye08, mich09}. 

\begin{deluxetable*}{ccccccc}\label{tab:fit}
\tabletypesize{\scriptsize}
\tablecolumns{7}
\tablewidth{0pt}
\tablecaption{Best-fit Results in UV--Optical--IR SED Fit \label{tab:fit}}
\tablehead{
\colhead{$z$} & \colhead{$M_{\star}$} & \colhead{$A_V$} & \colhead{Age} & \colhead{$\tau$} & \colhead{$Z$} & \colhead{$\chi^2$} \\
              & ($\MO$)               & (mag)           & (Myr)         & (Myr)            & ($\ZO$)       & \\
(1) &(2) &(3) &(4) &(5) &(6) &(7) 
} 
\startdata
$1.87^{+0.15}_{-0.07}$ &
$2.5^{+2.2}_{-0.3} \times 10^{11}$ &
$3.0^{+0.3}_{-1.0}$ &
$720^{+1880}_{-210}$ &
20 &
1  &
13.4 
\enddata
\tablecomments{
The errors are 68\% confidence intervals. 
(1) photometric redshift; 
(2) stellar mass; 
(3) $V$-band attenuation; 
(4) age; 
(5) star formation timescale; 
(6) metallicity; 
(7) $\chi^2$ value
}
\end{deluxetable*}

\subsection{Submm -- Radio}\label{sec:dust_sed}

\begin{figure}
\epsscale{1.15}
\plotone{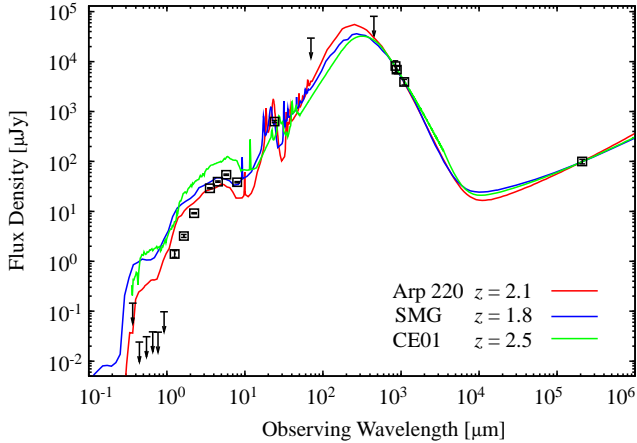}
\caption{
	Best-fit SEDs for three different models obtained in submm--radio SED fit. 
	The SEDs are the average SED of 76 SMGs \citep{mich09} at $z=1.8$, Arp~220 \citep{silv98} at $z=2.1$, and a starburst template of \cite{char01} at $z=2.5$. 
	Photometry data from UV to radio are overplotted. 
}
\label{fig:sed}
\end{figure}

Photometry at 850~$\micron$, 880~$\micron$, 1100~$\micron$, and 21~cm is used for submm-radio SED fits. 
The MIPS photometry is also used to check for consistency. 
We use the following starburst SED models: 
Arp~220 \citep{silv98}, 
the average SED of 76 SMGs with spectroscopic redshifts \citep{mich09}, 
and 105 SED models of \cite{char01}. 
We perform minimized $\chi^2$ fits with free parameters of redshifts and flux scaling factors. 
We find the best-fit redshifts of 
$z=2.1^{+0.5}_{-0.4}$ for Arp~220 ($\chi^2=0.3$), 
$z=1.8^{+0.5}_{-0.3}$ for the average SMG ($\chi^2=0.4$), 
and $z=2.5^{+0.6}_{-0.4}$ for an SED of \cite{char01} ($\chi^2=0.09$). 
The errors are 99\% confidence intervals. 
Note that there are other templates in the library of \cite{char01} that fit the photometry data with similar $\chi^2$ values in the redshift range of $z \sim 1$--3. 
The derived redshifts are consistent with the result in the UV-optical-IR SED fit in the previous section. 
The IR luminosities are estimated to be $L_{\rm IR} = (7$--$26) \times 10^{12}$~$\LO$ from the intrinsic IR luminosities of SED templates multiplied by scaling factors for the best-fit SEDs. 
The best-fit SED models are shown in Figure~$\ref{fig:sed}$ with photometry data from UV to radio. 
All of the SEDs overestimate the UV--NIR photometry, suggesting that stellar emission is heavily attenuated by dust. 
This is consistent with the large extinction ($A_V = 3.0$) derived from the UV--optical--IR SED fit. 
Finally, we note that the SMG sample in \citet{mich09} is biased against optically faint objects. 
It is clearly important to take into account a heavily obscured SMG like SXDF~850.6 (or e.g., GOODS 850-5 in \citealt{wang09}, AzTEC1 in Tamura et al. in prep.) to properly understand the overall picture of SMGs.

\section{Discussion}

\subsection{AGN Contribution}\label{sec:agn}

The rest-frame 1.6~$\micron$ bump is clearly seen at 5.8~$\micron$ (Figure \ref{fig:stellar_sed}), suggesting that the NIR emission detected in the IRAC bands is star-formation dominated \citep[e.g.,][]{weed06, farr08}. 
The flatter spectral slope in the IRAC emission compared to AGN populations favours a star formation as a dominant heating source \citep{yun08, ivis04}. 
This is also supported by the fact that SXDF~850.6 appears close to the starburst model tracks in the $S_{8.0}/S_{4.5}$ vs. $S_{24}/S_{8.0}$ color-color diagram of \cite{ivis04}. 
These facts suggest that SXDF~850.6 favours star formation as a dominant heating source and the physical quantities derived in the SED fit are less affected by the AGN. 

The 3$\sigma$ upper limit on the rest-frame X-ray luminosity derived from 0.2--12~keV flux is $1.3 \times 10^{44}$ erg~s$^{-1}$ (assuming $z=1.87$ and an effective photon index of $\Gamma=1.8$). 
Since this value is not properly corrected for hydrogen attenuation, we do not constrain the AGN contribution from the X-ray luminosity.

\subsection{Dust Mass and Molecular Gas Mass}\label{sec:mass}
Assuming the observed 880~$\micron$ flux is dominated by thermal dust emission, the dust mass can be derived as 
$M_d=S_{\rm{obs}}D_L^2/[(1+z)\kappa_d(\nu_{\rm{rest}})B(\nu_{\rm{rest}}, T_d)]$ \citep[e.g.,][]{hugh97}, 
where $S_{\rm{obs}}$ is the observed flux density, $\nu_{\rm{rest}}$ is the rest-frame frequency, $\kappa_d(\nu_{\rm{rest}})$ is the dust mass absorption coefficient, $T_d$ is the dust temperature, and $B(\nu_{\rm{rest}}, T_d)$ is the Planck blackbody function. 
We assume that the absorption coefficient varies as $\kappa_d \propto \nu^{\beta}$ and $\beta$ lies between 1 and 2 \citep[e.g,][]{hild83}. 
We adopt 
$\kappa_d(125\ \mu m) = 2.64 \pm 0.29$~m$^2$~kg$^{-1}$, the average value of various studies \citep{dunn03}, 
$T_d = 30$--50~K, and $\beta = 1.5$, the typical values for SMGs \citep[e.g.,][]{kova06, pope06, copp08, mich09}. 
The dust mass is estimated to be $M_d$ = (4--9) $\times 10^8$~$\MO$ for SXDF~850.6 at $z=1.87$. 
This is consistent with previous work on SMGs \citep[e.g.,][]{kova06, copp08, mich09}. 

By adopting a gas-to-dust mass ratio of 54, which is an average value for SMGs in \cite{kova06}, the molecular gas mass is $M_{\rm gas}$ = (2--5) $\times 10^{10}$~$\MO$. 

\subsection{Star Formation Activity and Nature of SXDF~850.6}

The UV--IR SED exhibits quiescent star forming activity dominated by old stellar components. 
The current SFR estimated from the SED fit is approximately zero. 
This is because the large part of the stellar mass was formed at the early phase of star formation and SFR decreased with time following exponential decay ($\tau = 20$~Myr for the best-fit SED). 
On the other hand, the IR luminosity of the best-fit submm--radio SED provides $\rm{SFR_{IR}}$ = 1300--4500~$\MO\ \rm{yr}^{-1}$ \citep{kenn98}. 
The SFR derived from 1.4~GHz radio emission is 1400~$\MO\ \rm{yr}^{-1}$ following the equation of \citet{bell03}, and 1100~$\MO\ \rm{yr}^{-1}$ following \citet{yun02} (with a spectral index of $-0.8$ and a radio--FIR normalization factor of 1). 

\begin{figure}
\epsscale{1.15}
\plotone{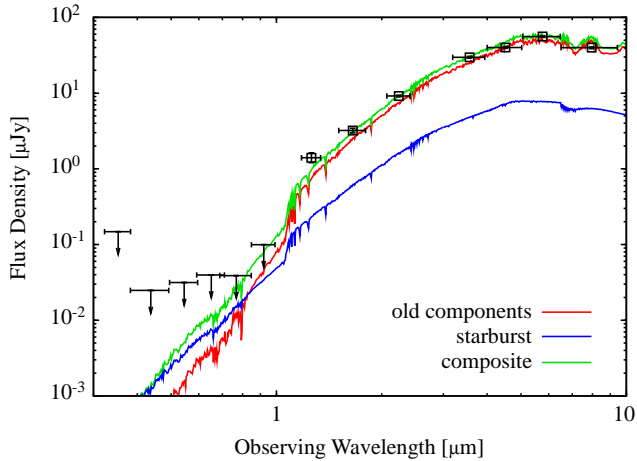}
\caption{
Synthetic SEDs of old stellar components (red), starburst components at the age of 100 Myr with an initial SFR of 3000~\MO\ yr$^{-1}$ (blue), and a sum of both (green) created with the GALAXEV library. 
Photometry data are also shown. 
}
\label{fig:composite_sed}
\end{figure}

To see whether the observational data allow the coexistence of the old stellar SED with quiescent star formation and the dusty starburst SED, we create synthetic SEDs at UV--IR wavelength using the GALAXEV library. 
The synthetic SEDs are composites of an SED with the same parameters as the best-fit SED which is dominated by old stellar population and starburst SEDs with different parameters. 
We find that the composite SEDs with plausible parameters for starburst SEDs (e.g., SFR $\sim$ 2000~$\MO\ \rm{yr}^{-1}$, $A_V = 3.0$, and age = 1~Myr) are consistent with the photometry data (Figure~\ref{fig:composite_sed}). 
Note that combining the SEDs increases the total stellar mass by only $\sim$1\%.

Combining these results allows us to infer the nature of SXDF~850.6: 
it is a mature system with a large fraction of old stellar components, and currently experiencing a vigorous dusty starburst. 
The coexistence of old stars and a current starburst in an SMG is suggested by \cite{wang09} from a detailed SED analysis. 
SXDF~850.6 has enough molecular gas mass as estimated in \S~\ref{sec:mass} to maintain intense star formation. 
Such significant star formation is likely caused by major mergers \citep[e.g.,][]{grev05, tacc06, tacc08, nara09}. 
If the star formation continues with SFR $\sim$ a few $10^3$~$\MO\ \rm{yr}^{-1}$, the gas consumption time is $\sim$ a few 10~Myr. 
This scenario is well described by hydrodynamic simulations of \citet{nara09} in which a major merger with a $\sim$$10^{13}$~$\MO$ dark matter halo produces a peak flux of $\sim$7--8~mJy at 850~$\micron$ for a short duration ($<$50~Myr). 
We are observing the short-lived glow of SXDF~850.6 as a bright SMG. 
Even if all the molecular gas converts into stars, the total stellar mass will increase by at most 30\%. 
This supports an idea that the large fraction of stellar mass in SMGs is constructed before the current submm bright phase \citep[e.g.,][]{bory05, dye08, mich09}.

The specific star formation rate (SSFR; SFR per unit stellar mass, an estimator of current star-forming activity) of $\sim$5--20~Gyr$^{-1}$ places SXDF~850.6 on the high mass end of the correlation between stellar mass and SFR for $z \sim 2$ star-forming galaxies or above the correlation \citep[e.g.,][]{erb06b, redd06, dadd07}. 
If the large population of stars were constructed before the current starburst phase, the enhancement of star formation could be due to merging of massive star forming galaxies \citep[e.g., merging BzKs;][]{taka08}, 
and it is possible that this galaxy evolves into  a massive present-day elliptical through major merger events \citep[e.g.,][]{lill99, smai04}.

\section{Summary}
We conducted SMA 880~$\micron$ observations of a submillimeter galaxy, SXDF~850.6, detected with SCUBA in the Subaru/\textit{XMM-Newton} Deep Field, which has multiple radio and IR counterpart candidates. 
The precise astrometry with the SMA shows that the correct counterpart is the most distant radio source from the SCUBA coordinates. 
Although there is a lack of a corresponding source at the SMA position in deep images of $u$, $B$, $V$, $R$, $i'$, $z'$ bands, 
we find a counterpart at $J$, $H$, $K$, 3.6~$\micron$, 4.5~$\micron$, 5.8~$\micron$, 8.0~$\micron$, 24~$\micron$, 1100~$\micron$, and 21~cm. 
A detailed analysis of fitting a library of synthetic SEDs to the photometry from the $u$-band to 8.0~$\micron$ gives the best-fit parameters of 
$z = 1.87^{+0.15}_{-0.07}$, 
$M_{\star} = 2.5^{+2.2}_{-0.3} \times 10^{11}\ \MO$, 
$A_V = 3.0^{+0.3}_{-1.0}$~mag, 
and age = $720^{+1880}_{-210}$~Myr. 
The SED fit to the submm--radio photometry using the three different starburst SED models provides the consistent redshift of $z \sim 1.8$--2.5 and $L_{\rm IR} = (7$--$26) \times 10^{12}$~$\LO$. 
The IR color diagnostic suggests that the AGN contribution to the SED is small. 
The SFR derived from the IR luminosity is 1300--4500~$\MO\ \rm{yr}^{-1}$, and the SSFR is $\sim$$10$ Gyr$^{-1}$. 
If the intense starburst continues, the estimated molecular gas mass of (2--5)~$\times 10^{10}$~$\MO$ will last for a few Myr. 
Given these facts, it is suggested that SXDF~850.6 is a mature system that already contains a dominant fraction of its expected final stellar mass in an old stellar population, and which is also experiencing a significant starburst.
The enhancement of star formation could be due to major mergers.

\acknowledgments
We thank the anonymous referee for valuable comments.
B.~H.\ is financially supported by a Research Fellowship from the JSPS for Young Scientists. 
JSD acknowledges the support of the Royal Society through a Wolfson Research Merit award.


\end{document}